\documentclass[conference]{IEEEtran}
\IEEEoverridecommandlockouts
\usepackage{cite}
\usepackage{amsmath,amssymb,amsfonts}
\usepackage{algorithm}
\usepackage{algpseudocode}
\usepackage{graphicx}
\usepackage{subcaption}
\usepackage{makecell}
\usepackage{amsmath}
\usepackage{blkarray}
\usepackage{textcomp}
\usepackage{xcolor}
\usepackage{balance}

\makeatletter
\AtBeginDocument{\BA@colsep=5pt}
\makeatother

\usepackage[all]{background}

\usepackage{stackengine}
\setstackEOL{\\}
\setstackgap{L}{\normalbaselineskip}
\SetBgContents{\color{gray}{\tiny \Longstack{PREPRINT - accepted by 21st IEEE Interregional NEWCAS Conference 2023 (NEWCAS 2023)}}}
\SetBgPosition{3.5cm,1cm}
\SetBgOpacity{1.0}
\SetBgAngle{0}
\SetBgScale{1.8}

\def\BibTeX{{\rm B\kern-.05em{\sc i\kern-.025em b}\kern-.08em
    T\kern-.1667em\lower.7ex\hbox{E}\kern-.125emX}}
\begin{document}


\title{Integrated Architecture for Neural Networks and Security Primitives using RRAM Crossbar}

\author{\IEEEauthorblockN{Simranjeet Singh\IEEEauthorrefmark{1}\IEEEauthorrefmark{5}\textsuperscript{\textsection}, Furqan Zahoor \IEEEauthorrefmark{2}\textsuperscript{\textsection}, Gokulnath Rajendran\IEEEauthorrefmark{2},
Vikas Rana\IEEEauthorrefmark{5},\\
Sachin Patkar\IEEEauthorrefmark{1},
Anupam Chattopadhyay\IEEEauthorrefmark{2}, Farhad Merchant\IEEEauthorrefmark{3} \IEEEauthorblockA{\IEEEauthorrefmark{1}Indian Institute of Technology, Bombay,  \IEEEauthorrefmark{2}Nanyang Technological University, Singapore, \\
\IEEEauthorrefmark{5}Forschungszentrum Jülich GmbH, \IEEEauthorrefmark{3}Newcastle University, UK}}\{simranjeet, patkar\}@ee.iitb.ac.in, \{furqan.zahoor@, gokulnat002@e., anupam@\}ntu.edu.sg, \\
\{si.singh, v.rana\}@fz-juelich.de, farhad.merchant@newcastle.ac.uk \vspace{-3mm} }

\maketitle

\begingroup\renewcommand\thefootnote{\textsection}
\footnotetext{Equal contribution}
\endgroup
\begin{abstract}
This paper proposes an architecture that integrates neural networks (NNs) and hardware security modules using a single resistive random access memory (RRAM) crossbar. The proposed architecture enables using a single crossbar to implement NN, true random number generator (TRNG), and physical unclonable function (PUF) applications while exploiting the multi-state storage characteristic of the RRAM crossbar for the vector-matrix multiplication operation required for the implementation of NN. The TRNG is implemented by utilizing the crossbar's variation in device switching thresholds to generate random bits. The PUF is implemented using the same crossbar initialized as an entropy source for the TRNG. Additionally, the weights locking concept is introduced to enhance the security of NNs by preventing unauthorized access to the NN weights. The proposed architecture provides flexibility to configure the RRAM device in multiple modes to suit different applications. It shows promise in achieving a more efficient and compact design for the hardware implementation of NNs and security primitives.

\end{abstract}

\begin{IEEEkeywords}
TRNG, PUF, NN, RRAM, Memristors, Hardware Security
\end{IEEEkeywords}

\section{Introduction}
\label{sec:intro}
The Internet of things (IoT) era has led to a significant increase in data exchange between processors and memory, which results in high power consumption. It ultimately degrades the system's performance~\cite{chen2022}. The von Neumann architecture, which separates processing and memory units, often suffers from data access latency due to the large volume of data movement~\cite{long2018}. To overcome these limitations, various novel computing paradigms are being investigated. In-memory computing, which performs calculations entirely within the computer memory, has gained significant attraction as a potential solution~\cite{ielmini2018, zidan2018, Staudigl2021}. 

Additionally, it is necessary to authenticate IoT devices on the network to ensure data security and protection by maintaining their integrity, confidentiality, and availability, thus preventing any malicious attacks or unauthorized access. Physical unclonable function (PUF) circuits are becoming popular in IoT due to their ability to generate unique and unpredictable responses to challenges. This makes them highly useful for hardware security, such as device authentication and key generation, and for implementing security protocols ranging from device attestation to data encryption. Several circuits have been proposed for realizing in-memory computing architectures using resistive random access memory (RRAM) devices to implement various techniques, including PUF~\cite{nili2018,john2021,singh2022}, neuromorphic neurons~\cite{huang2021,boybat2018} and digital gates~\cite{reuben2017,jeong2016,balatti2015}. RRAM is being considered as a potential candidate to address various drawbacks of conventional complementary metal oxide semiconductor (CMOS)-based architectures~\cite{mittal2014}. The relatively small size of RRAM devices also makes it highly feasible to integrate computing circuits and memory, thus realizing efficient architectures for learning algorithms, hardware security modules, and neural network (NN) applications~\cite{soudry2015}.

 \begin{figure}
    \centering
    \includegraphics[width=0.9\linewidth]{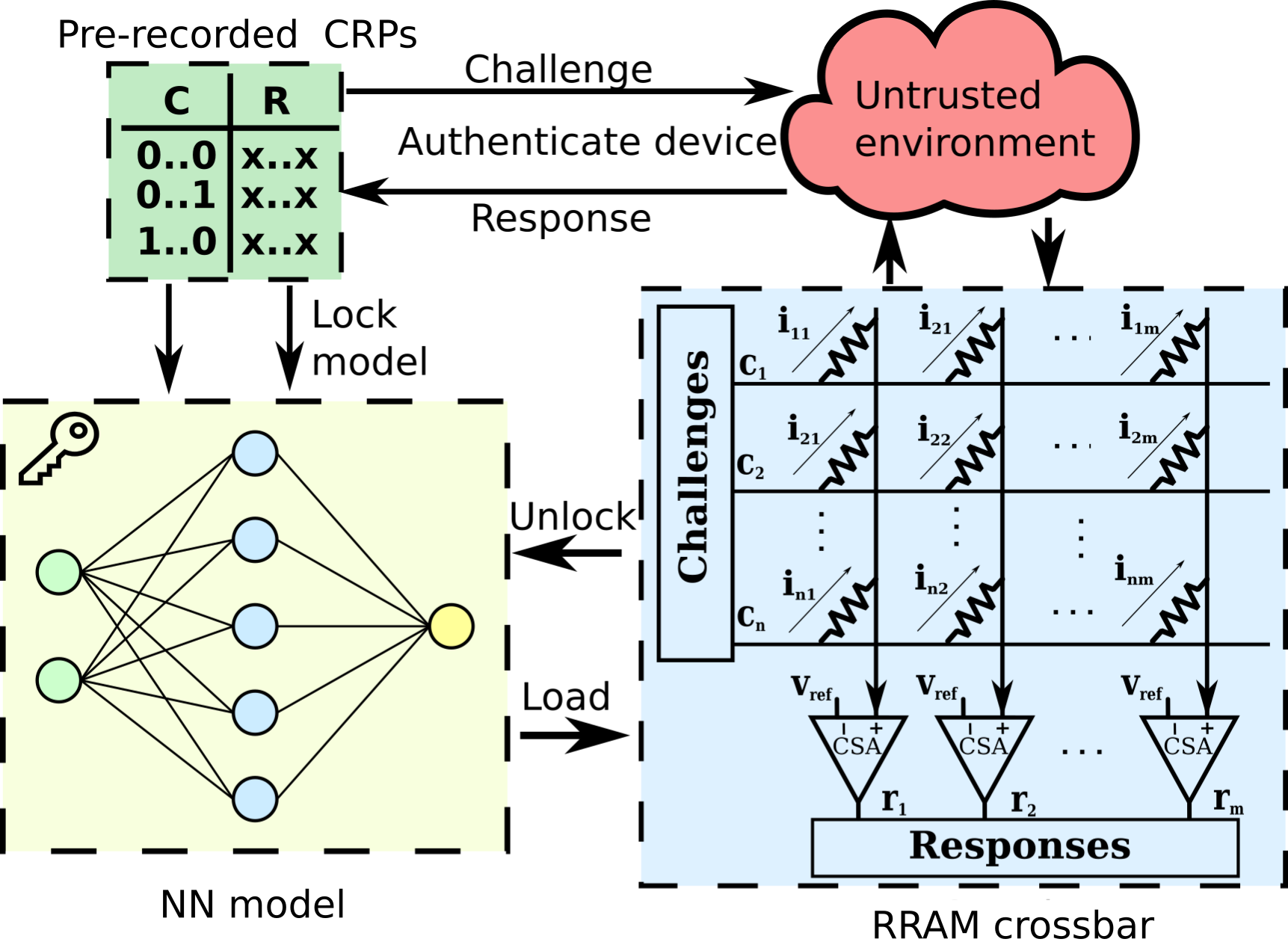}
    \caption{The NN locking using the embedded hardware security primitives by exploiting the variations of RRAM cells.}
    \vspace{-4mm}
    \label{fig:into}
\end{figure}
RRAM has been extensively studied for main memory and in-memory computing architectures, but their stochastic nature and intrinsic variation in switching parameters have hindered their widespread adoption as next-generation memories~\cite{zahoor2020}. However, this uncertain behavior is desirable for designing hardware security primitives~\cite{pang2017}. Another commonly investigated hardware encryption module is true random number generators (TRNGs), which generate a stream of random numbers by exploiting randomness in physical processes~\cite{govindaraj2018}. While CMOS-based TRNG designs have been proposed, they only provide limited security-specific properties, paving the way for TRNGs based on emerging technologies. Among these designs, RRAM-based TRNGs demonstrate desirable properties, primarily due to their low power operation, high density, and stochastic filament formation~\cite{yang2021}.

The protection of trained neural network (NN) models has become crucial to prevent unauthorized access, which can lead to the cloning of the model by adversaries. This study proposes a novel architecture to implement NN and hardware security modules on a single RRAM crossbar, allowing only authorized users with the correct device to use the locked NN model~\cite{chakraborty2020}. The proposed architecture focuses on protecting the intellectual property (IP) rights of deep NN models. Fig.~\ref{fig:into} illustrates the framework for locking the NN using the embedded security module. The major contributions of this work are as follows:

\begin{itemize}
    \item Integrating the NN, PUF, and TRNG on the RRAM crossbar array.
    \item Discussion on how the proposed crossbar architecture can be used for realizing NN weights locking.
    \item Lastly, the  methodology for implementing the NN, PUF, and TRNG on the same crossbar is validated.
\end{itemize}

The remainder of the paper is organized as follows: Section II details the architecture for implementing NN, TRNG, and PUF designs based on RRAM. Section III shows the NN weights-locking algorithm using the proposed architecture. Section IV discusses the results of integrating NN and hardware security primitives realized using the crossbar RRAM array. Section V concludes the~paper.

\section{Proposed Architecture}
\label{sec:proposed_arch}
\begin{figure}
    \centering
    \includegraphics[width=0.9\linewidth]{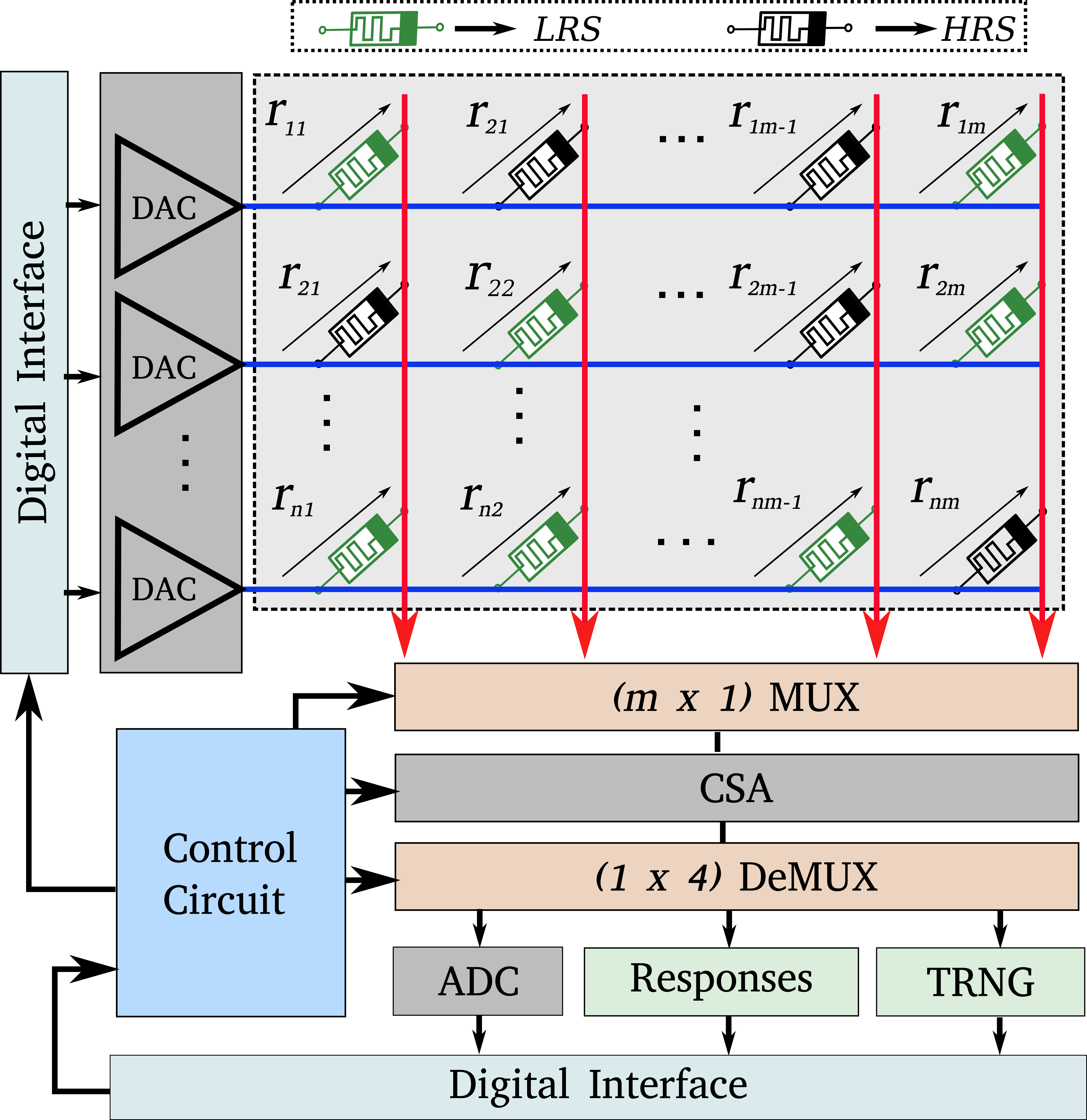}
    \caption{Proposed architecture for implementing NNs and hardware security primitives on the crossbar of RRAM.}
    \vspace{-4mm}
    \label{fig:arch}
\end{figure}

This section explains the architecture to integrate NN and hardware security modules using the RRAM crossbar. The proposed architecture is shown in Fig.~\ref{fig:arch}. The RRAM cells connected in a passive crossbar configuration are at the core of the proposed architecture. The same crossbar has been used to implement NN, PUF, and TRNG. 

\subsection{RRAM crossbar}
The RRAM device employed in this investigation is characterized by its ability to store multiple bits. Specifically, the device can be programmed into a low resistive state (LRS) and a high resistive state (HRS). Still, it can also store multiple resistive states between LRS and HRS, which is referred to as multi-state storage. Applying varying voltage pulses across the device can be configured as two- or multi-state devices. For the purpose of the vector-matrix multiplication (VMM) implementation in this study, the device is configured as a multi-state device, but it can also function in the two-state mode. The proposed architecture offers the flexibility to configure the device in multiple modes to suit applications such as VMM, TRNG, and PUF implementations. Next, we will discuss using RRAM as a core device to implement these applications on a single crossbar. 
\subsection{VMM implementation}
The RRAM crossbar performs the VMM operation, a critical function in implementing NNs. The weight matrix required for VMM is stored on a crossbar, with weights corresponding to each device's resistance or conductance state. These devices are set up to store multi-bit weights, and their rows are linked to a digital-to-analog converter (DAC). The input vector is then fed into the DACs, which generally convert the input into analog output voltage levels, as depicted in Fig.~\ref{fig:arch}. Following Ohm's current law, the applied input voltages produce a current through each device depending on the device's resistance or conductance value (weight). The current flowing through each device connected in a column is combined based on Kirchhoff's current law. Ultimately, the current values in the columns are utilized to carry out a multiplication and accumulation operation of the weights and input vector, which is nothing but a VMM operation in memory.

\begin{equation}
\begin{blockarray}{cc}
  & \\
\begin{block}{c(c)}
  v_1 & 1  \\
  v_2 & 2 \\ 
  v_2 & 2\\
  v_0 & 0\\
\end{block}
\end{blockarray}^{\raisebox{-1.5\baselineskip}{$T$}} 
\begin{blockarray}{ccccc}
 c_0 & c_1 & c_2 & c_3 &  \\
\begin{block}{(cccc)c}
  1 & 2 & 3 & 3  & r_0 \\
  0 & 3 & 0 & 1 & r_1 \\
  2 & 2 & 0 & 1 & r_2 \\
  3 & 2 & 2 & 1 & r_3 \\  
\end{block}
\end{blockarray} = 
\begin{blockarray}{cccc}
c_0 & c_1 & c_2 & c_3\\
\begin{block}{(cccc)}
  5 & 12 & 3 & 7  \\
\end{block}
\end{blockarray}
 \label{eq1}
\end{equation}

To illustrate, consider the VMM operation involving two-bit weights and two-bit input vectors. The input and weights matrix has been shown in Equation~\ref{eq1}. DAC maps the input vector to the voltage levels from $v_0$ to $v_3$. The weights matrix is stored as a resistance state on the crossbar (marked in Equation~\ref{eq1} using $r_n$ to $c_m$). The resistance state of the device in the crossbar represents the two-bit weights, while the input vector is applied to the DAC through a digital interface, as shown in Fig.~\ref{fig:arch}. The multiplication and accumulation results are obtained for each column, and the current sense amplifier for digital conversion amplifies the resulting current. In the example execution, the output matrix contains results greater than two bits, which need to be converted back to the digital domain. The analog-to-digital (ADC) resolution is determined by $\lceil \log_{2}(w \times m) \rceil$, where $w$ represents the weight bits (2 in this example) and $m$ means the number of devices in a single column (4 in this example).  

\subsection{TRNG}
To implement the TRNG on the RRAM crossbar, we have used the technique presented in~\cite{singh2023}, where the device-to-device (D2D) and cycle-to-cycle (C2C) variations on the crossbar have been used to generate the random switching in the crossbar. The switching threshold of each device on the RRAM crossbar is used to generate random bits. By applying a 50\% switching probability pulse to the crossbar, random devices switch their state to LRS, and the others remain in the HRS. Due to the crossbar variation, each device's switching threshold is different, resulting in random switching of the devices. In order to implement the TRNG in the proposed architecture, another terminal of the device must connect to GND, which the 1x4 DeMUX controls.

\subsection{PUF}
    A single RRAM crossbar is utilized in this proposed architecture to implement a TRNG and a PUF. The crossbar is first initialized to an entropy source based on the TRNGs algorithm, which provides a source of randomness to generate random bits. Challenges are then applied to the crossbar's rows, and the responses of the PUF are collected. The challenges are mapped to a read voltage pulse, which varies based on the input challenge. However, during PUF implementation, the crossbar is configured to two-state rather than multi-state switching.

    To collect the responses of the PUF, Kirchoff's current law is applied to the crossbar, which collects the current flowing through each device at the column lines. The input challenge and device variations influence this current. The sneak path affects the current in the crossbar, which contributes to the current at each column in a completely random manner. The analog current values are converted to boolean response bits at the output using a current sense amplifier (CSA). As the response bit can be either 0 or 1, ADC in the path has been bi-passed using  1x4 DeMUX. The digital interface can further use the collected responses to lock the weights matrix on the crossbar. The randomness and unpredictability of the PUF response make it suitable for use in secure authentication and key generation applications, and the incorporation of TRNG adds a layer of security.

\section{Weights locking}
\label{weight_lock}
Weights locking is a required method used to safeguard the intellectual property of NN models, particularly in scenarios where the model has been trained on sensitive data or where the model's performance is essential to business success. The proposed architecture integrates a PUF as a hardware security module in the RRAM crossbar. During training, the weights are encrypted using a unique key generated by the PUF. The architecture is configured to implement the PUF and generate the key to encrypt the weights. The encrypted weights and the challenge are then provided to the~user.

During the inference process, the architecture is reconfigured to implement the PUF, and the challenge provided with the encrypted weights is applied to generate the key. Next, the encrypted weights are loaded into the NN, and the key generated by the PUF is utilized to decrypt the weights. The decrypted weights are then used to make predictions.
\subsection{Locking}
The suggested design enables the NN to be secured within an RRAM crossbar. With the hardware security module situated on the same crossbar, a key can be generated via configuration in both the TRNG and PUF setups. The PUF-generated key can then be used to encrypt the weights to be stored in the crossbar. Algorithm~\ref{algo:lock} outlines the use of the proposed architecture for NN weights locking.

\begin{algorithm}
\caption{An algorithm for weights locking}\label{alg:cap}
\begin{algorithmic}[1]
\State Choose TRNG implementation
\State Apply 50\% probability  switching pulse
\State Choose PUF implementation
\State Apply challenge and collect the responses (key)
\State Store the CRPs on the server side and use a key to encrypt the weights. 
\State Send the encrypted weights to sharing platform with the challenge
\end{algorithmic}
\label{algo:lock}
\end{algorithm}

\subsection{Unlocking}
Once the weights have been encrypted, they can be transmitted to the user via any sharing platform. An identical challenge will be employed on the device's end to generate the required key. Algorithm~\ref{algo:unlock} specifies the decryption procedure. The device's inherent randomness is expressed via CRPs, which are unique to each device, which helps prevent attackers from using the same weights on a different hardware device.

\begin{algorithm}
\caption{An algorithm for unlocking the weights}\label{alg:cap}
\begin{algorithmic}[1]
\State Receive the encrypted weights
\State Choose TRNG implementation
\State Apply 50\% probability switching pulse
\State Choose PUF implementation
\State Apply challenge and generate the key
\State Choose the key to decrypt the weights
\State Store the decrypted weights on the same crossbar (overwrite the TRNG entropy)
\end{algorithmic}
\label{algo:unlock}
\end{algorithm}

In summary, this paper describes an architecture that integrates NNs and hardware security modules using passive RRAM cells connected in a crossbar structure. The RRAM device can store multiple resistive states between low and high resistive states, allowing it to function as a two-state or multi-state device. The proposed architecture offers flexibility to configure the device in multiple modes to suit different applications, such as VMM, TRNG, and PUF.

\section{Experimental Results}
\label{sec:Experimental_results}

For this study, an RRAM cell comprises a Pt/Ti/TiO$_x$/HfO$_2$/Pt material stack, demonstrating improved stability in terms of both electroforming voltage and thermal~\cite{Bengel2020}. This device can be configured into different configurations, such as binary and multi-state switching. The device is programmed into resistance by applying a voltage pulse with a specific duration and amplitude. The device exhibits resistance between $60 - 100K\Omega$ in HRS and $1.5 - 1.6K\Omega$ in LRS. However, in a multi-state configuration, there can be multiple states between HRS and LRS. The devices in the crossbar are utilized without any selector (passive) in series. Passive crossbars typically face the issue of sneak-path current, which can be utilized to design TRNGs and PUFs.

\subsection{Switching and Variations}
\begin{figure}%
\centering
\begin{subfigure}{0.43\columnwidth}
    \centering
    \includegraphics[width=\columnwidth]{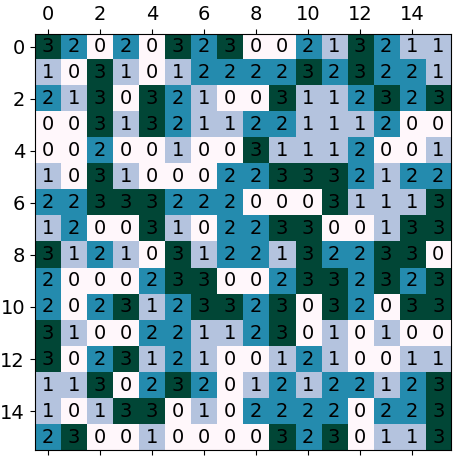}
     \caption{}
    \label{fig:chara}
\end{subfigure}\hfill%
\begin{subfigure}{.55\columnwidth}
    \centering
    \includegraphics[width=\columnwidth]{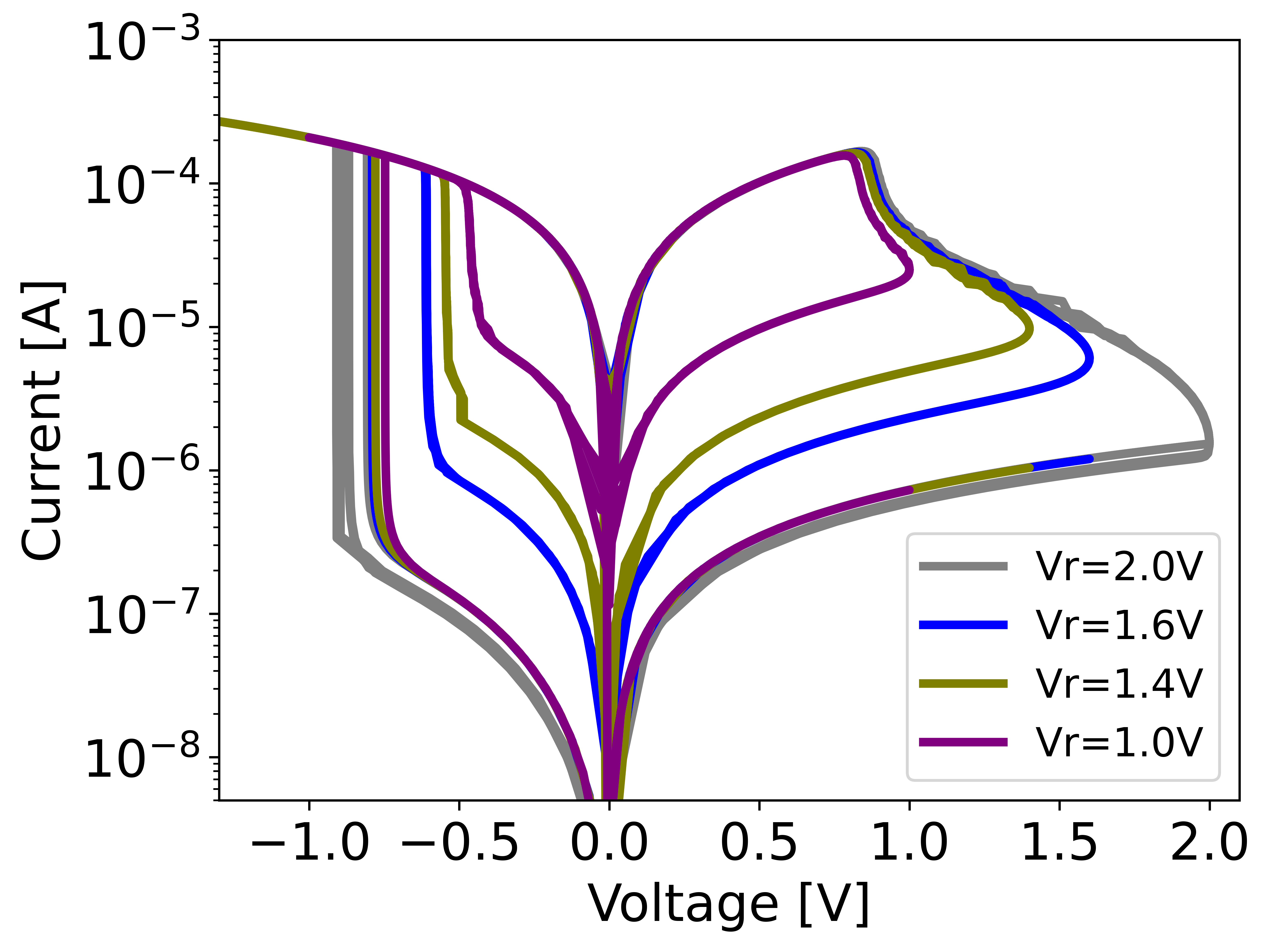}
    \caption{}
    \label{fig:charb}
\end{subfigure}\hfill%
\caption{The mapping of 2-bit weights to the crossbar is demonstrated in (a), while the multi-state behavior of devices for analog weight storage is illustrated in (b).\vspace{-4mm}}
\label{fig:char}
\end{figure}

In order to switch the device between binary states, a 150ns pulse of 2.0V with 10ns rise and the fall time is applied to switch the device to LRS (programming to 1), while a negative pulse of 2.0V is applied to switch the device to HRS (programming to 0). However, a gradual RESET method has been employed to achieve multi-state behavior. In this method, the device is initialized to LRS, and then an incremental pulse is applied to switch it to multiple states. The I-V curves of multi-state switching have been shown in Fig.~\ref{fig:char}.

The switching of the devices can be affected by variations in D2D and C2C parameters. These variations are influenced by manufacturing variations in device radius, device length, and oxygen ion concentration in the dielectric. As a result of these variations, the HRS resistance varies from 31$K\Omega$ to 155$K\Omega$ with an average of 65.56$K\Omega$, while the LRS resistance varies from 1.55$K\Omega$ to 1.67$K\Omega$ with an average of 1.64$K\Omega$. However, the HRS has a wide distribution; all HRS values are distinguishable from LRS. By carefully selecting the pulse to RESET the devices, they can be switched randomly from LRS to HRS or vice versa, which can be exploited to design the TRNGs.

\subsection{PUF properties}
We conducted extensive experiments to evaluate the reliability and performance of PUF on the proposed architecture. Our results demonstrate a reliability of 100\%, indicating that the CRPs generated by the proposed PUF are consistent and repeatable across multiple trials. Additionally, the uniqueness of the CRPs was found to be 47.78\%, meaning that the probability of generating the same CRP for two different devices is low. The uniformity of the CRPs was measured to be 49.79\%, indicating that the distribution of the CRPs is approximately uniform. Finally, the bit-aliasing was found to be 48.57\%, indicating that the probability of generating the same CRP for two different challenges is also low.

Importantly, our results show that the proposed PUF achieves these performance metrics without needing post-processing techniques, making it a practical and efficient hardware security primitive. Overall, our findings demonstrate the potential for using RRAM-based PUF designs in hardware security applications with high reliability and security characteristics.
\subsection{NN}
The crossbar can be utilized to map NN weights, similar to VMM applications. Fig.\ref{fig:chara} shows the mapping of 2-bit weights to the crossbar. To enable the multi-state behavior of a device, a gradual RESET method is employed, as illustrated in Fig.\ref{fig:charb}. We conducted a proof-of-concept by implementing VMM multiplication on a 16x16 crossbar, which is a critical operation for any NN implementation. The device's variations result in error accumulation at the input of the ADC. Nonetheless, this issue can be resolved through onboard fault-aware training of the required application.

\section{Conclusions}
\label{sec:conclusions}
In conclusion, this paper described a proposed architecture integrating NNs and hardware security modules using the RRAM crossbar as the core device. The proposed architecture enables a single crossbar to implement NN, TRNG, and PUF applications. The RRAM crossbar's multi-state storage characteristic is exploited to perform the vector-matrix multiplication operations required for NN implementation. The TRNG uses the crossbar's variation in device switching thresholds to generate random bits. The PUF is implemented using the same crossbar initialized as an entropy source for the TRNG. The proposed architecture provides flexibility to configure the RRAM device in multiple modes to suit different applications. This paper also presents the algorithms for NN weight locking. Overall, the proposed architecture shows promise in achieving a more efficient and compact design for the hardware implementation of NNs and security primitives.
\section*{Acknowledgments}
This work was supported in part by the Federal Ministry of Education and Research (BMBF, Germany) in the project NEUROTEC II under Project 16ME0398K, Project 16ME0399 and through Dr. Suhas Pai Donation Fund at IIT~Bombay.


\balance
\bibliographystyle{IEEEtran}
\bibliography{ref}

\end{document}